\documentclass[preprintnumbers,amsmath,amssymb,aps,prl]{revtex4-2}
\newcommand*{\eps}{{\rlap{\lower2ex\hbox{$\,\,\tilde{}$}}{\epsilon_{ijk}}}}
\newcommand*{\EPS}{{\rlap{\lower2ex\hbox{$\,\,\tilde{}$}}{\epsilon_{i'j'k'}}}}
\newcommand*{\lmq}{{\rlap{\lower2ex\hbox{$\,\,\tilde{}$}}{\epsilon_{lmq}}}}
\newcommand*{\jmq}{{\rlap{\lower2ex\hbox{$\,\,\tilde{}$}}{\epsilon_{jmq}}}}
\newcommand*{\jql}{{\rlap{\lower2ex\hbox{$\,\,\tilde{}$}}{\epsilon_{jql}}}}
\newcommand*{\jlm}{{\rlap{\lower2ex\hbox{$\,\,\tilde{}$}}{\epsilon_{jlm}}}}
\newcommand*{\imq}{{\rlap{\lower2ex\hbox{$\,\,\tilde{}$}}{\epsilon_{imq}}}}
\newcommand*{\iql}{{\rlap{\lower2ex\hbox{$\,\,\tilde{}$}}{\epsilon_{iql}}}}
\newcommand*{\ilm}{{\rlap{\lower2ex\hbox{$\,\,\tilde{}$}}{\epsilon_{ilm}}}}
\newcommand*{\lmn}{{\rlap{\lower2ex\hbox{$\,\,\tilde{}$}}{\epsilon_{lmn}}}}
\newcommand*{\abc}{{\rlap{\lower2ex\hbox{$\,\,\tilde{}$}}{\epsilon_{abc}}}}
\newcommand*{\N}{{\rlap{\lower2ex\hbox{$\,\,\tilde{}$}}{N}}}
\newcommand{\tN}{{\rlap{\lower2ex\hbox{$\,\,\tilde{}$}}{N}}}
\newcommand*{\tM}{{\rlap{\lower2ex\hbox{$\,\,\tilde{}$}}{M}}}
\newcommand*{\imn}{{\rlap{\lower2ex\hbox{$\,\,\tilde{}$}}{\epsilon_{imn}}}}

\begin{document}
\title{Quantization of constrained systems as Dirac first class versus second class: \\a toy model and its implications}

\author{Eyo Eyo Ita III}\email{ita@usna.edu}
\address{Physics Department, US Naval Academy. Annapolis, Maryland}
\author{Chopin Soo}\email{cpsoo@mail.ncku.edu.edu}
\address{Department of Physics, National Cheng Kung University, Taiwan}
\author{Abraham Tan}\email{mailfora.tan@gmail.com}
\address{Department of Physics, National Cheng Kung University, Taiwan}
\input amssym.def
\input amssym.tex

\bigskip\bigskip

\begin{abstract}
\par\indent
A toy model (suggested by Klauder) is analyzed from the perspective of First Class and Second Class Dirac constrained systems.
The comparison is made by turning a First Class into a Second Class system with the introduction of suitable auxiliary conditions. The links between Dirac's system of constraints,
the Faddeev-Popov canonical functional integral method and the Maskawa-Nakajima procedure to reduced phase space are explicitly illustrated.
The model reveals  stark contrasts and physically distinguishable results between First and Second class routes.
Physically relevant systems such as the relativistic point particle and electrodynamics are briefly recapped.
Besides its pedagogical value, the article also advocates the route of rendering First Class into Second Class systems prior to quantization.
Second Class systems lead to well-defined reduced phase space and physical observables; absence of inconsistencies in the closure of quantum constraint algebra; and consistent promotion of fundamental Dirac brackets to quantum commutators.
As First Class systems can be turned into well-defined Second Class ones, this has implications for the soundness of  ``Dirac quantization" of First Class constrained systems by simple promotion of Poisson, rather than Dirac brackets, to commutators without proceeding through Second Class procedures.
\end{abstract}

\maketitle

\section{1. Introduction}

The Hamiltonian formalism provides a natural starting point for the development of classical dynamics, and to quote Dirac ``For the purpose of setting up a quantum theory one must work from the Hamiltonian form" \cite{DIRAC}.  At the outset one has an even dimensional phase space $(q_i,p_i)$ of dimension $2N$ with symplectic
structure $\Omega=\sum_i^N{dq_i}\wedge{dp_i}$, and one can define functions on this phase space.  The Poisson bracket forms a Lie algebra on this space of functions
\begin{eqnarray}
\label{PBRACKETS}
\{A(\vec{q},\vec{p}),B(\vec{q},\vec{p})\}_{\text{P.B.}}=\sum_{i=1}^N\Bigl(\frac{\partial{A}}{\partial{q}_i}\frac{\partial{B}}{\partial{p}_i}
-\frac{\partial{B}}{\partial{q}_i}\frac{\partial{A}}{\partial{p}_i}\Bigr).
\end{eqnarray}
\noindent
The Hamiltonian $H(\vec{q},\vec{p},t)$ is a special phase space function which generates evolution with respect to a parameter $t$, known as time, of any quantity $g(t)$ via
\begin{eqnarray}
\label{GENERATE}
\dot{g}=\frac{\partial{g}}{\partial{t}}+\{q,H\}_{\text{P.B.}}.
\end{eqnarray}
For the choice $g(t)=(q_i(t),p_i(t))$, (\ref{GENERATE}) corresponds to the Hamilton's equations for the phase space variables.  The Hamilton's equations can be solved for the evolution of the system $(q_i(t),p_i(t))$ from initial
conditions $(q_i(0),p_i(0))$.  In its original formulation, quantization proceeds via the prescription of the promotion of dynamical variables to quantum operators $a\rightarrow\hat{a}$ and Poisson brackets to quantum commutators
\begin{eqnarray}
\label{PBRACKETS1}
[\hat{a},\hat{b}]=\hat{a}\hat{b}-\hat{b}\hat{a}=i\hbar\{a,b\}_{\text{P.B.}}.
\end{eqnarray}
\noindent
Modulo operator ordering ambiguities the Hamiltonian operator is constructed $\hat{H}(\vec{q},\vec{p},t)=H(\hat{\vec{q}},\hat{\vec{p}},t)$ and a wavefunction $\Psi[\vec{q},t]$ describing the state of the system can be defined based upon on the
configuration space $\vec{q}$ wherein the conjugate momenta act by differentiation.  Then one solves the Schrodinger equation for the
system $i\hbar\frac{\partial\Psi}{\partial{t}}=\hat{H}\Psi$ governing the unitary evolution of the state with respect to this time $t$ in respect of the quantum mechanical axioms.\par
\indent
There are various ambiguities and subtleties inherent in the construction of a quantum theory from its classical starting point (see, for instance, Ref.\cite{QUANTUM}).  For instance, by the Gronewald van Howe theorem it is in general not for all phase space functions $a(\vec{q},\vec{p})$ and $b(\vec{q},\vec{p})$ that one can consistently carry out (\ref{PBRACKETS1}), but rather only a limited subset\cite{QUANTUM}.  Also, the naive \textit{Poisson Brackets to Quantum Commutators} prescription presumes that there are no constraints upon the phase space variables.  This fails for constrained systems and the whole quantization procedure must be carefully modified in order to properly take constraints into account.  This will bring us to the main topic of this paper, namely totally constrained First Class versus Second Class systems and the interpretation of Dirac's formalism.\par
\indent
In Refs.\cite{DIRAC,LECTURES,LECTURES1} Dirac addresses the issue of constrained systems, classifying constraints $\Phi_I$ into either First Class or Second Class.  First Class constraints $\Phi_I$ have $\{\Phi_I,\Phi_J\}_{\text{P.B.}} = C^K_{IJ}\Phi_K$ while for Second
Class $\det(\{\Phi_I,\Phi_J\}_{\text{P.B.}}) \neq{0}$.  Dirac lays out the consistency conditions which must be satisfied for both types of constraints, both classically and upon quantization;
but these necessary conditions may not be sufficient to guarantee well-defined physical theories. This is a fundamentally important question regarding our universe, in particular in the case of totally constrained systems with weakly vanishing Hamiltonian, as in the usual canonical formulation of Einstein's theory \cite{DEWITT,ROVELLI,THIEMANN}.\par
\indent
Given a choice between First Class and Second Class constrained systems, which, if either,  should nature implement?  It is the proposition of this paper that Second Class are superior to First Class constrained systems in possessing a well-defined even-dimensional phase space and in the avoidance of inconsistencies in the constraint algebra with Dirac brackets. Indeed, Dirac allows for the possibility (on pages 33-36 of Ref.\cite{LECTURES}) that physical states be annihilated by First Class quantum constraints.  However, he does not specify whether the constraints should be quantized via Poisson brackets or otherwise, but only that $[{\hat\Phi}_I,{\hat\Phi}_J]={\hat C}^K_{IJ}{\hat\Phi}_K$ with structure functions ${\hat C}^K_{IJ}(\hat{\vec{q}},\hat{\vec{p}})$ appearing on the left as necessary to avoid inconsistency.  On the other hand, Second Class constraints must be solved and eliminated from the theory prior to quantization which proceeds by replacing Dirac brackets with commutators.\par
\indent
We will show that (\ref{PBRACKETS1}) fails for First Class constrained systems.  Some of the missing elements of First Class constrained Dirac quantization are that there is no clear phase space correspondence or direct link to perturbative quantum field theory via canonical functional integral formalism.  Indeed as we will show there cannot be a canonical functional integral without Faddeev--Popov determinants\cite{POPOV} and the Second Class Dirac scheme.  For example in the First Class constraint $q_1=0$ in a system with more than one variable, obviously $p_1=0$ is the proper subsidiary condition to convert to a Second Class system and hence the Dirac (not Poisson) bracket is zero.  Then one consistently promotes $q_1=0$, $p_1=0$ as operators without violation of canonical commutation relations (CCR).  On the other hand, the Poisson scheme will yield inconsistent $[q_1,p_1]=i\hbar$ and $p_1=\frac{\hbar}{i}\frac{d}{dq_1}$.  Suppose that the theory contains a Hamiltonian $H(p_1,p_2,\dots)$, then Poisson-Dirac First Class quantization will result in a nonsensical theory lacking a sensible classical limit. Dirac himself advocated that Second Class constrained systems be quantized via the Dirac bracket to commutator route. A First Class system does not have access to the Dirac bracket unless it is first turned into Second Class with supplementary conditions.\par
\indent
The proper relation amongst phase space variables is paramount.  A constraint $C$ implemented via a Lagrange multiplier $\lambda$ in the First Class approach can be deemed a phase space variable whose conjugate variable is $\dot{\lambda}$ and not $\lambda$.  So the natural question arises as to what is conjugate to $C$.  Since this is part of the redundant variables, then the conjugacy as per the Dirac--Faddeev--Popov scheme is that a subsidiary condition $\chi=0$ be taken to be equivalent to the new coordinate $\chi=Q=0$ and its conjugate arises via $\delta(P)\delta(Q)=\delta(C)\hbox{det}\{\chi,C\}_{P.B.}\delta(\chi)$.\par
\indent
This is in contrast with the naive Poisson--CCR--First Class Quantization wherein one promotes $\{q_i,p_j\}=\frac{1}{i\hbar}[\hat{q}_i,\hat{p}_j]$ in any system.  We will see that the correct prescription is to promote Dirac brackets $\{q_i,p_j\}_{\text{D.B.}}$ to quantum commutation relations.  In a system with Second Class constraints $\Phi_I$, the Dirac bracket is given by
\begin{eqnarray}
\label{FIRST10}
\{A,B\}_{\text{D.B.}}=\{A,B\}_{\text{P.B.}} -\{A,\Phi_I\}_{\text{P.B.}} (M^{-1})^{IJ}\{\Phi_J,B\}_{\text{P.B.}}
\end{eqnarray}
\noindent
with $M_{IJ}=\{\Phi_I,\Phi_J\}_{\text{P.B.}}$, wherein constraints and auxiliary conditions are lumped together into the same even dimensional phase subspace.  In other words, for a constrained system, quantum CCR \textit{cannot} be
determined \textit{before} Dirac brackets.  So given $\Phi(p_i,q_i)=0$ above, one cannot know what $\hat{\Phi}(\hat{p}_i,\hat{q}_i)=0$ is as a quantum equation.  This is not an operator-ordering issue.  One cannot even say $\hat{p}_i=\frac{\hbar}{i}\frac{d}{dq_i}$ because that would violate the Dirac brackets to CCR rule.  So $[\hat{p}_i,\hat{q}_j]$ can only be determined from promotion of Dirac brackets to commutators.\par
\indent
One of the main motivations of this paper is to highlight and address some of the difficulties associated with the treatment of General Relativity as a constrained system.  It is our view that the prescription whereby First Class constraints fix physical
states $\hat{\Phi}_I\vert\psi\rangle_{Phys}=0$ via quantization based upon Poisson brackets (see, for instance, Refs.\cite{THIEMANN,CONSTRAINT,CONSTRAINT1,CONSTRAINT2}) does not provide the correct interpretation of Dirac's work\cite{LECTURES}.  We will utilize an interesting toy model, for illustrative purposes.  This model will be known as the Klauder toy model for quantum gravity, due to John Klauder.  This consists of a two dimensional system with potential energy of a spring having a negative spring constant.  As we will see, this system will accentuate the differences between First Class and Second Class constrained systems, and will suggest the latter as a more viable approach.\par
\indent
In this paper we will argue that nature has a preference for Second Class constrained systems for several reasons.  Every system is either (1) Second Class, in which case it cannot be made First Class; (2) First Class, in which it can be made Second Class by introducing appropriate subsidiary conditions; or (3) partially First Class and partially Second Class in which case it can be made Second Class by (2).  By the Maskawa-Nakajima theorem\cite{MASKAWA}, Second Class systems inherit a natural phase space whereas First Class systems do not.  It should be noted that all variables satisfying Second Class constraints are observables  in the sense that, by construction and through the definition of the Dirac bracket, all variables have trivial Dirac brackets with all the constraints $\{A,\Phi_I\}_{\text{D.B.}} =0 \,\,\forall A, \Phi_I$, in particular constraints themselves also have trivial Dirac brackets $\{\Phi_I,\Phi_J\}_{\text{D.B.}} =0$ with each other.  They can thus be promoted to quantum observables; whereas the concept of Dirac observables weakly Poisson commuting with First Class constraints, and the constraint algebra closing weakly at the Poisson bracket level make local Dirac observables  extremely hard to find and the closure of the quantum constraint algebra a formidable challenge\cite{THIEMANN,ISHAM}.
\par
\indent
We will focus in this paper on the Klauder toy problem, highlighting the differences between First Class and Second Class constrained systems that it brings out, as well as its relation to General Relativity.

\section{2. Derivation of the Dirac bracket and reduced phase space}

We will now derive the Dirac bracket by explicit construction.  For a $2N$ dimensional phase space, given First Class constraints $\phi_i =0$ and auxilliary conditions $\chi_j=0$ with $i$ and $j$ ranging from 1 to $K < N$ such that:
\begin{equation}
\det(\{\chi_i,\phi_j\}_{\text{P.B.}}) \neq 0;
\end{equation}
we can set \(\{\chi_i\}\) as the first $K$ variables of a new set of general coordinates of the same phase space, we denote this by $^*$ superscript \(\chi_i=: q_i^*=0\).
Then $\det(\{\chi_i,\phi_j\}_{\text{P.B.}}) \neq 0$  translates into invertible $\{q_i^*,\phi_j\}_{\text{P.B.}}=\frac{\partial\phi_j}{\partial p_i^*}$.

Maskawa and Nakajima proved that for Second Class systems canonical  variables  exist  for  a  given  restricted  submanifold  that  the  submanifold is
specified  by  putting  those canonical variables equal  to zero. Thus we may decompose the total canonical phase space $\{ (q_I, p_I), I=1, ..., N \}$ into the sum of the reduced physical phase space  $\{ (q_r, p_r), r= K+1, ..., N\}$ and $\{ (q^*_i, p^*_i), i= 1, ..., K\}$;
and the constrained surface can be obtained by setting canonical $(q^*_i, p^*_i), i= 1, ..., K\}$ to zero (which is equivalent to satisfying $\phi_i =0, \chi_i =0, \forall i=1,.... K$).

The Poisson bracket between $A$ and $B$ for the reduced phase space would then be
\begin{equation}\label{eq:2}
\begin{split}
\{A,B\}_{\text{reduced}}&=\sum_{I=1}^N \left(\frac{\partial A}{\partial q_I}\frac{\partial B}{\partial p_I}-\frac{\partial A}{\partial p_I}\frac{\partial B}{\partial q_I}\right)-\sum_{j=1}^K \left(\frac{\partial A}{\partial q_j^*}\frac{\partial B}{\partial p_j^*}-\frac{\partial A}{\partial p_j^*}\frac{\partial B}{\partial q_j^*}\right)\\
&=\{A,B\}_{\text{P.B.}}-\sum_{j=1}^K \left(\{A,p_j^*\}_{\text{P.B.}}\{q_j^*,B\}_{\text{P.B.}}-\{q_j^*,A\}_{\text{P.B.}}\{B,p_j^*\}_{\text{P.B.}}\right)\\
&=\{A,B\}_{\text{P.B.}}-\sum_{j=1}^K \sum_{L=1}^N\Bigl[\Bigl(\frac{\partial A}{\partial q_L} \frac{\partial p^*_j}{\partial p_L} - \frac{\partial A}{\partial p_L}\frac{\partial p^*_j}{\partial q_L}\Bigr)\{q^*_j, B\}_{\text{P.B.}}
-\{q^*_j,A\}_{\text{P.B.}}\Bigl(\frac{\partial B}{\partial q_L} \frac{\partial p^*_j}{\partial p_L} - \frac{\partial B}{\partial p_L}\frac{\partial p^*_j}{\partial q_L}\Bigr)\Bigr]\\
&=\{A,B\}_{\text{P.B.}}-\sum_{j=1}^K \sum_{L=1}^N\Bigl[\Bigl(\frac{\partial A}{\partial q_L} \frac{\partial \phi_k}{\partial p_L} - \frac{\partial A}{\partial p_L}\frac{\partial \phi_k}{\partial q_L}\Bigl)\Bigl(\frac{\partial p^*_j}{\partial \phi_k}\Bigr)\{q^*_j, B\} _{\text{P.B.}}\\
&\qquad +\{A,q^*_j\}_{\text{P.B.}}\Bigl(\frac{\partial p^*_j}{\partial \phi_k}\Bigr)\Bigl(\frac{\partial B}{\partial q_L} \frac{\partial \phi_k}{\partial p_L} - \frac{\partial B}{\partial p_L}\frac{\partial \phi_k}{\partial q_L}\Bigr)\Bigr] \\
&=\{A,B\}_{\text{P.B.}}- (\{A,\phi_k\}_{\text{P.B.}} {\{\phi_k ,\chi_j\}}^{-1}\{\chi_j,B\}_{\text{P.B.}}+\{A,\chi_j\}_{\text{P.B.}} {\{\chi_j,\phi_k\}}^{-1}\{\phi_k,B\}_{\text{P.B.}}),\\
&=\{A,B\}_{\text{D.B.}}.
\end{split}
\end{equation}
Thus the Dirac bracket is equivalent to the Poisson bracket w.r.t. total phase space subtracted by the unwanted phase space $(q^*_i , p^*_i)$  eliminated by the $\chi_i$ and $\phi_i$.
The correct evolution of an operator in a constrained system should be w.r.t. to the physical reduced phase space implies the Dirac, rather than Poisson, bracket should be used to arrive at the right physical evolution. i.e.
$\dot A = \{A,H\}_{\text{D.B.}}$, wherein $H$ is the Hamiltonian of the system.\\

Interestingly, the following holds true:
\begin{equation}
\begin{split}
&\text{Let }A\vert_c:=A\rvert_{\chi=0,\phi=0}, \:B\vert_c:=B\rvert_{\chi=0,\phi=0}\\
\{A\rvert_c,B\rvert_c\}_{\text{P.B.}}&=\sum_{j=1}^K \left(\frac{\partial (A\vert_c)}{\partial q_j^*}\frac{\partial (B\vert_c)}{\partial p_j^*}-\frac{\partial (A\vert_c)}{\partial p_j^*}\frac{\partial (B\vert_c)}{\partial q_j^*}\right)+\sum_{r=K+1}^N \left(\frac{\partial (A\vert_c)}{\partial q_r}\frac{\partial (B\vert_c)}{\partial p_r}-\frac{\partial (A\vert_c)}{\partial p_r}\frac{\partial (B\vert_c)}{\partial q_r}\right)\\
&=0+\sum_{r=K+1}^N \left(\frac{\partial A}{\partial q_r}\frac{\partial B}{\partial p_r}-\frac{\partial A}{\partial p_r}\frac{\partial B}{\partial q_r}\right)\Bigg|_{\chi=0,\phi=0}\\
&=\{A,B\}_{\text{D.B.}}\rvert_{\chi=0,\phi=0},\\
\end{split}
\end{equation}
The first entry of the intermediate step vanishes because on the constrained surface $\delta q^*_i =0$, so $A\rvert_c = A(q_r, p_r)$  and  $B\rvert_c = B(q_r, p_r)$ have no dependence on $q^*_i$.
That the second entry is equivalent to the Dirac bracket is precisely a theorem of Maskawa and Nakajima: the Dirac bracket of a Second Class system is equal the Poisson bracket w.r.t. to the reduced variables.

\section{3. Sympletic analysis: Relativistic particle, and Maxwell theory}

The treatment of Second Class constrained systems resides in the existence of an invertible map between a subset of the starting phase space and the constraint and auxiliary condition that preserves the existence of an even-dimensional phase space.  This can also be seen using symplectic two forms.  We will start off with two known examples, highlighting various attributes that each brings to the table.\par
\indent
The first example is the relativistic point particle in Minkowski spacetime.  In the first class approach one starts from
action $S=\int\bigl(p_{\mu}\frac{dx^{\mu}}{d\tau \hfill}-\lambda{C}\bigr)d\tau$ wherein the mass shell condition $\frac{1}{2}\bigl(p^{\mu}p_{\mu}-m^2\bigr)=0=C$ becomes implemented as a constraint via Lagrange multiplier $\lambda$.  Naive application of the Dirac procedure wherein First Class quantum constraints annihilate physical states (upon promotion
$\hat{p}_{\mu}=-i\hbar\partial_{\mu}$) leads to the Klein--Gordon equation
\begin{eqnarray}
\label{KLEIN}
\hat{C}\psi
=\frac{1}{2}\Bigl(\frac{\partial^2}{\partial{t}^2}-\partial^2-\frac{m^2}{\hbar^2}\Bigr)\psi=0;\\
\psi(\vec{x},t)=\frac{1}{(2\pi)^3}\int{d}^3k\Bigl(A(\vec{k})e^{(i/\hbar)(\sqrt{k^2+m^2}t-\vec{k}\cdot\vec{x})}+
B(\vec{k})e^{-(i/\hbar)(\sqrt{k^2+m^2}t-\vec{k}\cdot\vec{x})}\Bigr)
\end{eqnarray}
\noindent
whose solution consists of forward and backward propagation corresponding to positive and negative energy modes.  This is inconsistent with the invariance of the state under transformations with respect to the parameter $\tau$ generated by the constraint through the promotion to commutators of Poisson brackets of
\begin{eqnarray}
\label{NAIVE}
\frac{dx^i}{d\tau}=\{x^i,\lambda{C}\}_{\text{P.B.}}=\lambda{p}^i;~~\frac{dp_i}{d\tau}=\{p_i,\lambda{C}\}_{\text{P.B.}}=0
\end{eqnarray}
\noindent
since (1) the polarization of the state is not preserved on configuration space $x^i$ as $\psi(x^i,\int\lambda{d}\tau)=\psi(x^i+p^i\int\lambda{d}\tau,\int\lambda{d}\tau)$ on account of the constraint's being quadratic in momentum, as well as (2) the arbitrariness in $\lambda(\tau)$ stemming from reparametrization invariance.  The evolution is fictitious, having nothing to do with physical time evolution.  The analogue of this is the Wheeler--deWitt equation for gravity, wherein one has the problem of time\cite{CONSTRAINT1,ISHAM}.\par
\indent
Conversely, in the Second Class approach one proceeds upon the premise of preservation of an even-dimensional phase space.  With starting sympletic two form
\begin{eqnarray}
\label{POINT}
\Omega={dx^{\mu}}\wedge{dp_{\mu}}=\Omega_{Phys}+\Omega_{constr.}=-{dx_i}\wedge{dp_i}+{dx^0}\wedge{dp_0}
\end{eqnarray}
\noindent
wherein $\Omega_{constr.}={dx^0}\wedge{dp_0}$, we choose the following constraint and auxiliary condition
\begin{eqnarray}
\label{POINT1}
C=\frac{1}{2}(p_{\mu}p^{\mu}-m^2)=\frac{1}{2}(p_0^2-p_ip_i-m^2);~~\xi=x^0-\tau.
\end{eqnarray}
\noindent
The physical intepretation is the emergence of a physical Hamiltonian $p_0=\sqrt{p_ip_i+m^2}$ and a time $x^0=\tau$ upon implementation. This is similar to the emergence of a reduced Hamiltonian in the full theory of gravitation\cite{SOOITAYU}.

One sees on the constraint surface that $\{\chi,C\}=p_0=\sqrt{p_ip_i+m^2}\neq{0}$, and that the constraint and auxilliary condition produce a symplectic two form ${d\chi}\wedge{d{C}}={dx^0}\wedge{\bigl(p_0dp_0-p_idp_i\bigr)}$, yielding the desired invertible map
\begin{eqnarray}
\label{POINT2}
{dx^0}\wedge{dp_0}=\frac{{d\chi}\wedge{dC}}{\{\chi,C\}}+\frac{p_i}{p_0}{dx^0}\wedge{dp_i}
\end{eqnarray}
\noindent
with a cross term ${dx^0}\wedge{dp_i}$ between $\Omega_{Phys}$ and $\Omega_{constr.}$.  Substituting (\ref{POINT2}) into (\ref{POINT}) we have for the physical part of the phase space, after subtracting out the constrained part
\begin{eqnarray}
\label{POINT3}
\Omega_{Phys}=\Omega-\frac{{d\chi}\wedge{dC}}{\{\chi,C\}}=-{\Bigl(dx^i-\frac{p_i}{p_0}dx^0\Bigr)}\wedge{dp_i}.
\end{eqnarray}
\noindent
The vanishing of this symplectic two form $\Omega_{Phys}=0$ follows from Hamilton's equation for $x^i$ from emergent $H_{Phys}=\sqrt{p_ip_i+m^2}$ generating evolution in time $t=\tau$, both arising from the constraints.  The same time $t$ governs evolution of the wavefunction $\psi(x^i,t)$ via Schrodinger equation $i\hbar\frac{\partial\psi}{\partial{t}}=\hat{H}\psi$ with solution
\begin{eqnarray}
\label{POINT4}
x^i(\tau)=x^i(0)+\frac{p^i\tau}{\sqrt{p_ip_i+m^2}};~~\psi(\vec{x},\tau)=
\hbox{exp}\Bigl[-i\tau\sqrt{-\partial^2+\frac{m^2}{\hbar^2}}\Bigr]\psi(\vec{x},0)
\end{eqnarray}
\noindent
corresponding to time evolution within a single branch of (\ref{KLEIN}).  From a physical point of view one must have positive energy solutions propagating forward in time via unitary evolution of quantum mechanics.  On the other hand, evolution via (\ref{NAIVE}) must be interpreted as a gauge transformation, which is unphysical.\par
\indent
The second relevant example will be the Maxwell theory with starting symplectic two form
\begin{eqnarray}
\label{MAXWELL}
\Omega=\int{d}^3x~{\delta{A}_i(x)}\wedge{\delta{E}_i(x)}=
\int{d}^3x~{\delta{A^T}_i(x)}\wedge{\delta{E^T}_i(x)}
+\int{d}^3x~{\delta{A^L}_i(x)}\wedge{\delta{E^L}_i(x)}
\end{eqnarray}
\noindent
with the association of $\Omega_{Phys}$ and $\Omega_{constr.}$ to the transverse and longitudinal parts, respectively, of the vector potential and electric field $(A_i,E_i)$.  We choose the constraint and auxiliary condition
\begin{eqnarray}
\label{MAXWELL1}
C=\partial_iE_i;~~\chi=\partial_iA_i
\end{eqnarray}
\noindent
whose interpretation is the imposition of transversality for physical fields.  The constraint $C=\partial_iE_i=0$ is the Gauss' law constraint.  One sees that for appropriately defined conditions, $\{\chi(x),C(y)\}=\partial^2\delta(x-y)\neq{0}$ and that the constraint and auxiliary condition imply
${\delta\chi(x)}\wedge{\delta{C}(x)}={\partial_i(\delta{A}_i(x))}\wedge{\partial_i(\delta{E}_i(x))}$ and that we have
for the longitudinal part of the phase space
\begin{eqnarray}
\label{MAXWELL2}
{\delta{A_i^L}(x)}\wedge{\delta{E_i^L}(x)}
=\frac{{\delta\chi(x)}\wedge{\delta{C}(x)}}{\{\chi,C\}}
=\frac{1}{\partial^2}{\partial_i(\delta{A}_i(x))}\wedge{\partial_j(\delta{A}_j(x))}.
\end{eqnarray}
\noindent
Substituting (\ref{MAXWELL2}) back into (\ref{MAXWELL}), we have for the transverse part of the phase space
\begin{eqnarray}
\label{MAXWELL3}
\Omega_{Phys}=
\int{d}^3x~{\delta{A^T}_i(x)}\wedge{\delta{E^T}_i(x)}
=\int{d}^3x~\Bigl(\delta_{ij}-\frac{\partial_i\partial_j}{\partial^2}\Bigr)
{\delta{A}_i(x)}\wedge{\delta{E}_j(x)}
\end{eqnarray}
\noindent
with transverse, albeit non-local, projector.  There is no Hamiltonian and no time, as neither the Gauss' law constraint nor auxiliary condition, unlike the previous example, involve a Hamiltonian or a time.  To obtain nontrivial time evolution, both classical and quantum mechanically, we must choose a Hamiltonian involving the transverse physical fields.  For the choice of Hamiltonian density $\frac{1}{2}\bigl(E^T_iE^T_i+\partial^2A^T_iA^T_i\bigr)$, when we have a Physical Hamiltonian and a physical time whose evolution is generated via Dirac brackets
\begin{eqnarray}
\label{MAXWELL4}
\dot{A}^T_i(x)=E^T_i(x);~~\dot{E}^T_i(x)=-\partial^2A^T_i(x);\\
i\hbar\frac{\partial\psi[A^T_i,t]}{\partial{t}}
=\int{d}^3x\frac{1}{2}\Bigl(-\hbar^2\frac{\delta^2}{\delta{A}^T_i(x)\delta{A}^T_i(x)}
+\partial^2{A^T(x)}A^T_i(x)\Bigr)\psi[A^T_i,t]
\end{eqnarray}
\noindent
with a Schrodinger equation governing the evolution of the state with respect to the same time $t$.  In this case the analogue of (\ref{NAIVE}), the Gauss' law constraint 
\begin{eqnarray}
\label{MAXWELL5}
\partial_i\frac{\delta\psi}{\delta{A}_i(x)}=0\longrightarrow\psi(A_i,t)=\psi(A_i+\partial_i\lambda,t)
\end{eqnarray}
\noindent
can be correctly interpreted as invariance of the state under gauge transformations
$\delta{A}_i(x)=\{A_i(x),\int{d}^3y\lambda(y)\partial_iE_i(y)\}_{\text{P.B.}}$ upon promotion to quantum commutators on account of the constraint's being linear in momentum.  Additionally, there is no conflict with time evolution in this case, which is not determined through the constraints.

\section{4. The klauder toy problem: Canonical equivalence of the second class approach}

We will now move on to the Klauder toy problem for quantum gravity, which will have various features in common with the previous two examples.  Let us consider a 2-dimensional starting phase space in polar coordinates $(q_i,p_i)=(r,\varphi,p_r,p_{\varphi})$.  In accordance with the procedure for handling Second Class constraints we need to impose, for each constraint $C={0}$ on this phase space, a suitable auxilliary condition $\chi={0}$.  For the constraint $ C = \frac{1}{2} (p^2_r+\frac{L^2_\varphi }{r^2} -\alpha^2 r^2)=0$ we can choose the auxilliary condition $\chi=rp_r-k=0$ respectively.  Here, $k$ is constant with respect to the phase space variables but may in general have explicit time dependence.  $L_\varphi := p_\varphi$ is actually the generator of ``z-axis rotations" for all the variables; moreover, $L_\varphi$ commutes with $C$ and also with $\chi$  since they are both rotationally invariant.  Note, other than at the point $(p, r) = (0,0)$, that the criterion
\begin{eqnarray}
\label{CRITERION}
\{\chi, C\}_{\text{P.B.}} = p_r^2+\frac{p^2_{\varphi}}{r^2}+\alpha^2r^2=p^2 +\alpha^2r^2 \neq 0
\end{eqnarray}
\nonumber\
for Dirac's Second Class system is met, where $p$ is the total momentum of the system $p^2=p_r^2+\frac{p^2_{\varphi}}{r^2}$.\par
\indent
To see the implications for the canonical phase space path integral, it is instructive to examine the system in the formalism of differential forms.  The symplectic two form on the total unconstrained phase space is given by
\begin{eqnarray}
\label{DIFFERENTIAL}
\Omega=\sum_{I=1}^2{\delta{q}_I}\wedge{\delta{p}_I}
={\delta\varphi}\wedge{\delta{p}_{\varphi}}+{\delta{r}}\wedge{\delta{p}_r}.
\end{eqnarray}
\noindent
The variations of the constraint and auxilliary condition are given by
\begin{eqnarray}
\label{CONDITIONALITY}
\delta{C}=p_r\delta{p}_r+\frac{p_{\varphi}}{r^2}\delta{p}_{\varphi}-\Bigl(\frac{p^2_{\varphi}}{r^3}+\alpha^2r\Bigr)\delta{r};~~
\delta\chi=r\delta{p}_r+p_r\delta{r}.
\end{eqnarray}
\noindent
Thus the constraint and auxilliary condition furnish a symplectic structure
\begin{eqnarray}
\label{STRUCTURE}
{\delta\chi}\wedge{\delta{C}}
=\frac{p_{\varphi}}{r^2}{\Bigl(r\delta{p}_r+p_r\delta{r}\Bigr)}\wedge{\delta{p}_{\varphi}}
+\Bigl(p_r^2+\frac{p^2_{\varphi}}{r^2}+\alpha^2r^2\Bigr){\delta{r}}\wedge{\delta{p_r}}\\
=\frac{p_{\varphi}}{r^2}{\delta(rp_r)}\wedge{\delta{p}_{\varphi}}+\{\chi,C\}{\delta{r}}\wedge{\delta{p_r}}
\end{eqnarray}
\noindent
One sees that the transformation $(r,p_r)\rightarrow(\chi,C)$ saturates the constrained part of the phase space, whereupon the conditions $C=0$, $\chi=0$ can be imposed as strong equalities.  Then $rp_r=k$ implies $\delta(rp_r)=\delta{k}=0$
and $\{\chi,C\}_{\chi=0,C=0}=2\alpha^2r^2=\frac{2(k^2+p^2_{\varphi})}{r^2}$ and subject to the constraint and auxilliary condition we have
\begin{eqnarray}
\label{CONDITION2}
{\delta\chi}\wedge{\delta{C}}\biggl\vert_{\chi=0,C=0}=
\{\chi,C\}{\delta{r}}\wedge{\delta{p_r}}\longrightarrow
{\delta{r}}\wedge{\delta{p_r}}=\frac{{\delta\chi}\wedge{\delta{C}}}{2\alpha^2r^2},
\end{eqnarray}
\noindent
with a one-to-one correspondence to the relevant part of the canonical phase space implementing these conditions.  In fact, there is a strong analogy to the Maxwell problem in that the angular phase space $(\varphi,p_{\varphi})$ play the analogous role to the transverse field $(A_i^T,E_i^T)$ as does the radial phase space $(r,p_)$ to the longitudinal $(A_i^L,E_i^L)$ as the phase space variables will form an orthogonal decomposition with respect to some suitable inner product.  The canonical functional integral measure with constraint and auxilliary condition is
\begin{eqnarray}
\int \,dr\, dp_r\, d\varphi\, dp_\varphi \,\, \det(\{\chi, C\}_{\text{P.B.}})\delta(\chi)\delta(C)
~~\hbox{exp}\Big[\frac{i}{\hbar}\int\Bigl(p_rdr+p_{\varphi}d\varphi -Hdt\Bigr)\Bigr]\\
=\int \,dr\, dp_r\, d\varphi\, dp_\varphi \,\, (p^2 +\alpha^2r^2)\delta(\chi=rp_r-k)\delta(C)
~~\hbox{exp}\Big[\frac{i}{\hbar}\int\Bigl(p_rdr+p_{\varphi}d\varphi-Hdt\Bigr)\Bigr].
\end{eqnarray}
\noindent
Subject to the constraint and auxilliary condition there is a one-to-one mapping $(r,p_r)\rightarrow(\chi,\phi)$, with contribution to the phase space measure given by

\begin{displaymath}
{dr}dp_r=\hbox{det}
\left(\begin{array}{cc}
\frac{\partial\chi}{\partial{r}} & \frac{\partial\chi}{\partial{p}_r}\\
\frac{\partial{C}}{\partial{r}} & \frac{\partial{C}}{\partial{p}_r}\\
\end{array}\right)^{-1}
d\chi{dC}=\hbox{det}
\left(\begin{array}{cc}
p_r & r\\
-\Bigl(\frac{p^2_{\varphi}}{r^3}+\alpha^2r\Bigr) & p_r\\
\end{array}\right)^{-1}
d\chi{dC}=\frac{d\chi{d}C}{p^2+\alpha^2r^2}
\end{displaymath}
\noindent
Substituting into the measure we have
\begin{eqnarray}
\label{MEASURE}
\int\frac{d\chi{d}C}{p^2+\alpha^2r^2}{d}\varphi{d}p_{\varphi}(p^2+\alpha^2r^2)\delta(\chi)\delta(C)~~
\hbox{exp}\Big[\frac{i}{\hbar}\int\Bigl(p_rdr+p_{\varphi}d\varphi-Hdt\Bigr)\Bigr]\\
=\int{d\varphi}dp_{\varphi}drdp_r\delta(p_r-p_r^{*})\delta(r-r^*)~~
\hbox{exp}\Big[\frac{i}{\hbar}\int\Bigl(p_rdr+p_{\varphi}d\varphi-Hdt\Bigr)\Bigr]\\
=\int{d\varphi}dp_{\varphi}e^{(i/\hbar)\int{p^*}dr^*}e^{(i/\hbar)\int{p}_{\varphi}d\varphi}e^{-(i/\hbar)\int{H(r^*,p^*,\varphi,p_{\varphi})}dt}
\end{eqnarray}
\noindent
The correctness of the measure is guaranteed by canonical equivalence
\begin{eqnarray}
\label{MEASURE1}
\int{d}\chi{d}C\delta(\chi)\delta(C)=\int{d}rdp_r\delta(r-r^*)\delta(p_r-p_r^{*});\\
r^{*}=\Bigl(\frac{k^2+p^2_{\varphi}}{\alpha^2}\Bigr)^{1/4};~~p^{*}_r=\frac{k}{r^*}
=k\Bigl(\frac{k^2+p^2_{\varphi}}{\alpha^2}\Bigr)^{-1/4}.
\end{eqnarray}
\noindent
wherein the reduced phase space can be identified as $(\varphi,p_{\varphi})$.  It can be verified the Dirac brackets are as follows:
\begin{eqnarray}
\{\varphi, p_\varphi\}_{\text{D.B.}}&=& \{\varphi, p_\varphi\}_{\text{P.B.}} - \{\varphi, \chi\}_{\text{P.B.}}\frac{-1}{\{\chi, C\}_{\text{P.B.}}}\{C, p_\varphi\}_{\text{P.B.}} -\{\varphi, C\}_{P.B}\frac{+1}{\{\chi, C\}_{\text{P.B.}}}\{\chi,p_\varphi\}_{\text{P.B.}} \\
&=&\{\phi, p_\phi\}_{\text{P.B.}} =1,
\end{eqnarray}
since $\{\chi, p_\varphi=L_\varphi\}_{\text{P.B.}}=0 = \{C, p_\varphi \}_{\text{P.B.}}$ due to rotational invariance under $L_\varphi$ as noted above.
The Dirac bracket of reduced resultant pair $(\varphi, p_\varphi)$ equals the Poisson bracket, signifying that the pair is indeed canonical even with the constraint solved and auxilliary condition imposed.
On the other hand,
\begin{eqnarray}
\{r, p_r\}_{\text{D.B.}}&=& \{r, p_r\}_{\text{P.B.}} - \{r, \chi\}_{\text{P.B.}}\frac{-1}{\{\chi, C\}_{\text{P.B.}}}\{C, p_r\}_{\text{P.B.}} -\{r, C\}_{P.B}\frac{+1}{\{\chi, C\}_{\text{P.B.}}}\{\chi, p_r\}_{\text{P.B.}} \\
&=&\{r, p_r\}_{\text{P.B.}} - (r)\frac{-1}{p^2+\alpha^2 r^2}(-\frac{p^2_\varphi}{r^3} -\alpha^2 r) -(p_r)\frac{1}{p^2+\alpha^2 r^2}(p_r) \\
&=& 1-\frac{1}{p^2+\alpha^2 r^2}(\frac{p^2_\varphi}{r^2} +\alpha^2 r^2+p^2_r) =0,
\end{eqnarray}
\noindent
indicating the pair $(r,p_r)$ is no longer canonical!\par
\indent
The remaining Dirac brackets are given by
\begin{eqnarray}
\label{REMAINING}
\{p_r,p_{\varphi}\}_{\text{D.B.}}=\{p_r,p_{\varphi}\}_{\text{P.B.}} - \{p_r, \chi\}_{\text{P.B.}}\frac{-1}{\{\chi, C\}_{\text{P.B.}}}\{C, p_{\varphi}\}_{\text{P.B.}} -\{p_r, C\}_{P.B}\frac{+1}{\{\chi, C\}_{\text{P.B.}}}\{\chi, p_{\varphi}\}_{\text{P.B.}}\\
-(-p_r)\Bigl(\frac{-1}{p^2+\alpha^2 r^2}\Bigr)(0)-\Bigl(\frac{p_{\varphi}^2}{r^3}+\alpha^2r\Bigr)\Bigl(\frac{1}{p^2+\alpha^2 r^2}\Bigr)(0)=0,
\end{eqnarray}
\noindent
with cross-brackets
\begin{eqnarray}
\label{REMAINING1}
\{p_{\varphi},r\}_{\text{D.B.}}=\{p_{\varphi},r\}_{\text{P.B.}} - \{p_{\varphi}, \chi\}_{\text{P.B.}}\frac{-1}{\{\chi, C\}_{\text{P.B.}}}\{C, r\}_{\text{P.B.}} -\{p_{\varphi}, C\}_{P.B}\frac{+1}{\{\chi, C\}_{\text{P.B.}}}\{\chi, r\}_{\text{P.B.}}\\
=-(0)\Bigl(\frac{-1}{p^2+\alpha^2 r^2}\Bigr)(-p_r)-(0)\Bigl(\frac{1}{p^2+\alpha^2 r^2}\Bigr)(-r)=0\\
\{p_r,\varphi\}_{\text{D.B.}}=\{p_r,\varphi\}_{\text{P.B.}} - \{p_r, \chi\}_{\text{P.B.}}\frac{-1}{\{\chi, C\}_{\text{P.B.}}}\{C, \varphi\}_{\text{P.B.}} -\{p_r, C\}_{P.B}\frac{+1}{\{\chi, C\}_{\text{P.B.}}}\{\chi, \varphi\}_{\text{P.B.}}\\
=0-(-p_r)\Bigl(\frac{-1}{p^2+\alpha^2 r^2}\Bigr)(-\frac{p_{\varphi}}{r^2})-\Bigl(\frac{p_{\varphi}^2}{r^3}-\alpha^2r\Bigr)\Bigl(\frac{1}{p^2+\alpha^2 r^2}\Bigr)(0)=\frac{p_rp_{\varphi}}{r^2(p^2+\alpha^2r^2)},
\end{eqnarray}
\noindent
The vanishing of any Dirac bracket with $p_{\varphi}$ is consistent with rotational invariance.  And finally, we have
\begin{eqnarray}
\label{REMAINING2}
\{r,\varphi\}_{\text{D.B.}}=\{r,\varphi\}_{\text{P.B.}} - \{r, \chi\}_{\text{P.B.}}\frac{-1}{\{\chi, C\}_{\text{P.B.}}}\{C, \varphi\}_{\text{P.B.}} -\{r, C\}_{P.B}\frac{+1}{\{\chi, C\}_{\text{P.B.}}}\{\chi, \varphi\}_{\text{P.B.}}\\
=0-(r)\Bigl(\frac{-1}{p^2+\alpha^2 r^2}\Bigr)\Bigl(-\frac{p_{\varphi}}{r^2}\Bigr)
-(p_r)\Bigl(\frac{1}{p^2+\alpha^2 r^2}\Bigr)(0)
=-\frac{p_{\varphi}}{r(p^2+\alpha^2r^2)}
\end{eqnarray}
\noindent

\section{5. Dynamics on the reduced phase space}

In summary of the previous results the Dirac brackets of the variables give the following relations:
\begin{equation}
\begin{split}
\{r,p_r\}_{\text{D.B.}}  &  =0\\
\{r,p_\varphi\}_{\text{D.B.}} & =0\\
\{r,\varphi\}_{\text{D.B.}} & =-\frac{r p_\varphi}{p_\varphi^2+ r^2 p_r^2+\alpha^2 r^4 }\\
\{\varphi,p_r\}_{\text{D.B.}} & =-\frac{p_r p_\varphi}{p_\varphi^2+ r^2 p_r^2+\alpha^2 r^4 }\\
\{\varphi,p_\varphi\}_{\text{D.B.}} & =1\\
\{p_r,p_\varphi\}_{\text{D.B.}} & =0
\end{split}
\end{equation}
\noindent
Solving for the variables from the \(\chi,C\) restrictions, we get the following:
\begin{equation}
\begin{split}
& r^2 p_r^2+\alpha^2 r^4 = p_\varphi^2+2 k^2 \\
& r= \sqrt[4]{\frac{k^2+p_\varphi^2}{ \alpha^2}};~~p_r=\frac{k}{r}\\
\{r,\varphi\}_{\text{D.B.}} & =-\frac{r p_\varphi}{2(p_\varphi^2+k^2)}=-\frac{p_\varphi}{2(p_\varphi^2+k^2)} \sqrt[4]{\frac{k^2+p_\varphi^2}{ \alpha^2}}
\:\xrightarrow{k \rightarrow 0}\:-\frac{1}{2 p_\varphi^2} \sqrt[4]{\frac{p_\varphi^2}{m \alpha^2}}\\
\{\varphi,p_r\}_{\text{D.B.}} & =-\frac{k p_\varphi}{2 r (p_\varphi^2+k^2)}=-\frac{k p_\varphi}{2(p_\varphi^2+k^2)} \sqrt[4]{\frac{ \alpha^2}{k^2+p_\varphi^2}}
\:\xrightarrow{k \rightarrow 0}\: 0
\end{split}
\end{equation}
Had we substituted the solution of $(r, p_r)$ in terms of $(\varphi, p_{\varphi})$ and computed the resultant P.B., we would have obtained the same results. This is a consequence of a theorem we proved earlier.\par
\indent
Since all phase space functions $f(q_I,p_I)$ Dirac-commute with the constraints $\{f,C\}_{\text{D.B.}}=\{f,\chi\}_{\text{D.B.}}=0$ for all $f$, then all phase space functions, in Second Class constrained systems, are said to be Dirac observables.  As the system has been completely reduced without an emerging Hamiltonian, a physical Hamiltonian $H_{Phys}$ needs to be prescribed to discuss dynamics and also classical orbits.  This is exactly analogous to the situation in the Maxwell problem considered earlier.  To wit we introduce the general 2-dimensional non-relativistic Hamiltonian with
radial potential which preserves the invariance under rotations generated by $p_{\varphi}$ which was present in $C$ and $\chi$.
  For any radial potential \(V(r)\), the Hamiltonian \(H=\frac{p_r^2}{2}+\frac{p_\varphi^2}{2r^2}+V(r)\) gives the following relation:
\begin{eqnarray}
\label{DYNAMICAL}
H_{Phys}=\frac{p_r^2}{2}+\frac{p_\varphi^2}{2 r^2}-\frac{1}{2}\alpha^2r^2+U(r)=C+U(r)=\frac{1}{2}\alpha^2r^2+V(r)
\end{eqnarray}
\noindent
wherein $C=0$ has been imposed as a strong equality.  In Second Class constrained systems, constraints and auxiliary conditions may be \textit{a-priori} imposed upon the system prior to computation of evolution via Dirac brackets.  This is in contrast to First Class systems, wherein constraints can be implemented only after
Poisson brackets have been evaluated.  In the former case a genuine physical Hamiltonian $H_{Phys}$ serves as the generator of evolution of the reduced variables with respect to a parameter $t$, which defines the time
\begin{equation}
\begin{split}
\{r,H_{Phys}\}_{\text{D.B.}} & =\frac{\partial{H}_{Phys}}{\partial{r}}\{r,r\}_{\text{D.B.}}= 0\\
\{p_r,H_{Phys}\}_{\text{D.B.}} & =\frac{\partial{H}_{Phys}}{\partial{r}}\{p_r,r\}_{\text{D.B.}}=0\\
\{\varphi,H_{Phys}\}_{\text{D.B.}} & =\frac{\partial{H}_{Phys}}{\partial{r}}\{\varphi,r\}_{\text{D.B.}}=\frac{r p_\varphi U'(r)}{p_\varphi^2+ r^2 p_r^2+\alpha^2 r^4 }\\
\{p_\varphi,H_{Phys}\}_{\text{D.B.}} & =\frac{\partial{H}_{Phys}}{\partial{r}}\{p_{\varphi},r\}_{\text{D.B.}}=0,\\
\text{where }U'(r)=\frac{dU}{dr}.\\
\end{split}
\end{equation}
\noindent
Dirac brackets, as do Poisson brackets, obey the Leibniz rule from calculus.  Substituting the above relation with \(\chi=0,\phi=0\), we get:
\begin{eqnarray}
\label{WEGET}
\{\varphi,H_{Phys}\}_{\text{D.B.}}=-\frac{p_{\varphi}U'(r^{*})}{2\alpha^2{r^*}^3};~~r^*=\Bigl(\frac{k^2+p^2_{\varphi}}{\alpha^2}\Bigr)^{1/4}
\end{eqnarray}
Since the Dirac bracket of a variable with a Hamiltonian is the time-evolution of that variable under that Hamiltonian, we have
\begin{eqnarray}
\label{WEGET11}
\dot{r}=\dot{p}_r=\dot{p}_{\varphi}=0;~~\dot{\varphi}=-\frac{p_{\varphi}U'(r^{*})}{2\alpha^2{r^*}^3};\\
r(t)=\Bigl(\frac{k^2+p^2_{\varphi}}{\alpha^2}\Bigr)^{1/4}=r^*;~~
p_r(t)=k\Bigl(\frac{k^2+p^2_{\varphi}}{\alpha^2}\Bigr)^{-1/4}=\frac{k}{r^*};~~
\varphi(t)=\varphi(0)-\frac{p_{\varphi}U'(r^{*})}{2\alpha^2{r^*}^3}t;~~p_{\varphi}(t)=p_{\varphi}
\end{eqnarray}
\noindent
It is interesting to note that the system will only have circular orbits regardless of the value of $k$ for time-independent $k$, since \(r\) is constant in time since $p_{\varphi}$ is conserved as it Dirac commutes with the Hamiltonian.  With respect to observational applications, the results thus derived should be true just for experiments or observations that fulfill the auxiliary condition $p_r=\frac{k}{r}$.

\section{6. Quantization}

Canonical quantization proceeds by promotion of all dynamical variables to operators $A\rightarrow\hat{A}$ and by promoting all Dirac, {\it not} Poisson, brackets to quantum
commutators $\{A,B\}_{\text{D.B.}}=\frac{1}{i\hbar}[\hat{A},\hat{B}]$.  Even Hamilton's equation for phase space function $O(\vec{q},\vec{p},t)$ must follow this rule
\begin{eqnarray} 
\label{HEISENBERG}
\frac{dO}{dt}=\frac{\partial{O}}{\partial{t}}+\{O,H_{Phys}\}_{D.B.}\longrightarrow
\hat{\Bigl(\frac{dO}{dt}\Bigr)}=\frac{\partial\hat{O}}{\partial{t}}+\frac{1}{i\hbar}[\hat{O},\hat{H}_{Phys}],
\end{eqnarray} 
\noindent 
yielding Heisenberg's equation of motion for quantum operator $\hat{O}$. For the fundamental dynamical variables, we have
\begin{eqnarray}
\label{COMMUTATOR}
[\hat{r},\hat{p}_r]=[\hat{r},\hat{p}_{\varphi}]=[\hat{p}_r,\hat{p}_{\varphi}]=0;~~[\hat{\varphi},\hat{p}_{\varphi}]=i\hbar
\end{eqnarray}
\noindent
Surprisingly, values of $(r,p_r)$, which is albeit a non-commuting pair at the Poisson bracket level, can be determined with infinite precision quantum mechanically because $[{\hat r}, {\hat p}_r]= i\hbar\{r, p_r\}_{\text{D.B.}}=0$.  For the remaining commutators, note that all variables may \textit{a-priori} be reduced by the constraint and auxiliary condition prior to quantization
\begin{eqnarray}
\label{COMMUTATOR11}
[\hat{\varphi},\hat{r}]=\frac{i\hbar}{2\sqrt{\alpha}}\hat{p}_{\varphi}(\hat{p}^2_{\varphi}+k^2)^{-3/4};~~
[\hat{p}_r,\hat{\varphi}]=\frac{i\sqrt{\alpha}k\hbar}{2}\hat{p}_{\varphi}(\hat{p}^2_{\varphi}+k^2)^{-5/4}
\end{eqnarray}
\noindent
As all commutators have been reduced to dependence upon a single fundamental variable $p_{\varphi}$, there are no operator ordering ambiguities.  Quantization of the theory gives eigenstates which obey ${\hat L}_\varphi \Psi_m(\varphi) =(m\hbar )\Psi_m(\varphi) $, and all CCR have a well-defined action on eigenstates of $p_{\varphi}$.  The physical Hamiltonian operator is also given by $\hat{H}_{Phys}=U(r^{*}(\hat{p}_{\varphi}))$.\par
\indent
We require, as a consistency condition, that the time $t$ of (\ref{HEISENBERG}) govern unitary evolution of the quantum state $\psi$ via the Schrodinger equation.  For the Klauder toy problem we have
\begin{eqnarray}
\label{COMMUTATOR2}
i\hbar\frac{\partial\psi}{\partial{t}}=\hat{H}_{Phys}\psi;~~
\psi(\varphi,t)=\hbox{exp}\Bigl[-\frac{itU(\hat{p}_{\varphi})}{\hbar}\Bigr]\psi(\varphi,0)
\end{eqnarray}
\noindent
The configuration space of the system is the unit circle $S^1$.  Let us choose the initial wavefunction, thus polarized, as a normalizable superposition of eigenstates of angular momentum
\begin{eqnarray}
\label{COMMUTATOR3}
\psi(\varphi,0)=\sum_{m\in{Z}}c_me^{im\varphi};\\
\langle\psi\vert\psi\rangle=
\frac{1}{2\pi}\int^{2\pi}_0d\varphi~\psi^{*}(\varphi,0)\psi(\varphi,0)=\sum_{m\in{Z}}\vert{c}_m\vert^2=1
\end{eqnarray}
\noindent
wherein $c_m$ are the Fourier coefficients.  For the totally constrained case the state remains as in \eqref{COMMUTATOR3} for all time, which is a Heisenberg picture state.  But under physical Hamiltonian $H_{Phys}=U(r)$ we have $U(\hat{r})\vert{m}\rangle=U_m\vert{m}\rangle$, and the state evolves via
\begin{eqnarray}
\label{COMMUTATOR4}
\psi(\varphi,t)=\sum_{m\in{Z}}c_me^{im\varphi} e^{-(i/\hbar)U_mt}
\end{eqnarray}
\noindent
which preservation of its normalizability.  Expectation values go through as well
\begin{eqnarray}
\label{COMMUTATOR5}
\Bigl<\hat{A}\Bigr>=\frac{1}{2\pi}\int^{2\pi}_0d\varphi~\psi^{*}(\varphi,t)\hat{A}(\varphi,\hat{p}_{\varphi})\psi(\varphi,t).
\end{eqnarray}
\noindent
We can compare expectation values of Schrodinger picture operators in the state (\ref{COMMUTATOR4}) with the classical evolution via (\ref{WEGET}) and (\ref{WEGET11}).  We have the following results
\begin{eqnarray}
\label{EXPECTATION}
\Bigl<\hat{r}\Bigr>=\sum_{m\in{Z}}\vert{c}_m\vert^2\Bigl(\frac{k^2+(m\hbar)^2}{\alpha^2}\Bigr)^{1/4};\\
\Bigl<\hat{p}_r\Bigr>=\sum_{m\in{Z}}\vert{c}_m\vert^2k\Bigl(\frac{k^2+(m\hbar)^2}{\alpha^2}\Bigr)^{-1/4};~~
\Bigl<\hat{p}_{\varphi}\Bigr>=\hbar\sum_{m\in{Z}}m\vert{c}_m\vert^2
\end{eqnarray}
\noindent
The variables which in the classical theory are time-independent, yield quantum operators which do not evolve from their stationary states.  This time independence is a consequence of $k$ having no explicit time dependence, which results in a time independent physical Hamiltonian.  For the angle $\varphi$ we have
\begin{eqnarray}
\label{EXPECTATION1}
\Bigl<\hat{\varphi}\Bigr>=\frac{1}{2\pi}\sum_{m,n}e^{(i/\hbar)(U_m-U_n)t}c^{*}_mc_n\int^{2\pi}_0d\varphi~\varphi~e^{i(n-m)\varphi}\\
=\pi+i\sum_{n\neq{m}}\frac{e^{(i/\hbar)(U_m-U_n)t}c^{*}_mc_n}{n-m}
\end{eqnarray}
\noindent
with nontrivial time evolution.  An interesting question is whether canonical transformations on the phase space commute with quantization.  While, as noted, there are no ordering ambiguities in polar coordinates, ones sees that ordering ambiguities exist in Cartesian coordinates
\begin{eqnarray}
\label{TRANSFORMATON}
x=r\hbox{cos}\varphi;~~y=r\hbox{sin}\varphi;\\
p_x=p_r\hbox{cos}\varphi-\frac{p_{\varphi}}{r}\hbox{sin}\varphi;~~
p_y=p_r\hbox{sin}\varphi+\frac{p_{\varphi}}{r}\hbox{cos}\varphi.
\end{eqnarray}
\noindent
on account of the noncommutativity of $\varphi$ and $p_{\varphi}$.  Nevertheless, expectation values can be computed for a given ordering, using the polar coordinate quantization.  We have
\begin{eqnarray}
\label{TRANSFORMATION1}
\langle\hat{x}+i\hat{y}\rangle=\sum_{m,n}r^{*}_nc^{*}_mc_ne^{(i/\hbar)(U_m-U_n)t}\delta_{m,n+1};~~
\langle\hat{p}_x+i\hat{p}_y\rangle=\sum_{m,n}\frac{1}{r^{*}_n}c^{*}_mc_n(k+i(m\hbar))e^{(i/\hbar)(U_m-U_n)t}\delta_{m,n+1}.
\end{eqnarray}
\noindent 
where $r^{*}_n=\alpha^{-1/2}(k^2+(n\hbar)^2)^{-1/4}$.

\section{7. First class constraints approach}

The Klauder toy problem in the first class approach with Lagrange multiplier $\lambda$, can well be illustrated in Cartesian coordinates via starting action (using the summation convention)
\begin{eqnarray}
\label{FIRST}
S=\int\Bigl(p_i\dot{q}_i-\frac{\lambda}{2}(p_ip_i-\alpha^2q_iq_i)\Bigr)dt.
\end{eqnarray}
\noindent
Canonical equivalence with its representation in polar coordinates furnishes unconstrained canonical Poisson brackets $\{q_i,q_j\}_{\text{P.B.}}=\{p_i,p_j\}_{\text{P.B.}}=0$ and $\{q_i,p_j\}_{\text{P.B.}}=\delta_{ij}$.  The corresponding Hamilton's equations read
\begin{eqnarray}
\label{FIRST1}
\dot{q}_i=\lambda{p}_i;~~\dot{p}_i=\lambda\alpha^2q_i;~~C=p_ip_i-\alpha^2q_iq_i=0.
\end{eqnarray}
\noindent
The third equation is the constraint $C=0$ which follows from the equation of motion for the Lagrange multiplier $\lambda$, which at this stage can be an arbitrary function of time.  It can also be viewed, within the context of the Dirac procedure, as a secondary constraint arising from preservation of the primary constraint $p_{\lambda}=0$, where $p_{\lambda}$ is the conjugate momentum to the Lagrange multiplier $\lambda$.  Then we have $\dot{p}_{\lambda}=C=0$.\par
\indent
The solution can be written down by inspection
\begin{eqnarray}
\label{WEHAVEIT}
q_i(T)=q_i(0)\hbox{cosh}(\alpha{T})+\frac{p_i(0)}{\alpha}\hbox{sinh}(\alpha{T});~~
p_i(T)=p_i(0)\hbox{cosh}(\alpha{T})+\alpha{q}_i(0)\hbox{sinh}(\alpha{T}),
\end{eqnarray}
\noindent
with $T=\int^t_0dt'\lambda(t')$r.  But $T$ is arbitrary on account of the arbitrariness of $\lambda$, so (\ref{WEHAVEIT}) is not physical time evolution.  Time reparametrization invariance  implies the state $(q_i(t),p_i(t))$ is related to initial state $(q_i(0),p_i(0))$ by some sort of group action.  The group is abelian, noncompact and preserves an invariant hyperbolooid, similarly to preservation of the light cone under Lorentz transformations in special relativity.  Note, throughout its evolution, the system respects the constraint $C(q_i(t),p_i(t))=C(q_i(0),p_i(0))$.\par
\indent
That the constraint is preserved under evolution via Poisson brackets can also be seen directly from the Hamilton's equations (\ref{FIRST1})
\begin{eqnarray}
\label{PRESERVE}
\frac{d}{dt}(p_ip_i-\alpha^2q_iq_i)=2p_i\dot{p}_i-2\alpha^2q_i\dot{q}_i=2p_i(\lambda\alpha^2q_i)-2\alpha^2q_i(\lambda{p}_i)=0.
\end{eqnarray}
\noindent
No further constraints are generated and the system is First Class.  Indeed, on page 21 of \cite{LECTURES} Dirac states (and we paraphrase) \textit{Primary First Class constraints are generating functions of infinitesimal contact transformations, which lead to changes in the q's and p's that do not affect the physical state.}  Within the context of First Class systems, one must interpret the auxiliary condition as a gauge-fixing condition.  Let
\begin{eqnarray}
\label{GAUGEFIXING}
\chi=\vec{q}\cdot\vec{p}-k(t)=0,
\end{eqnarray}
\noindent
where $k=k(t)$ has explicit time dependence, the reason for which will be clear momentarily.  In a gauge-fixed First Class system, we require the gauge-fixing condition to be preserved for all time.  So
\begin{eqnarray}
\label{MANDATE}
\frac{d\chi}{dt}=
\frac{d(q_ip_i)-k(t)}{dt}=\dot{q}_ip_i+q_i\dot{p}_i-\dot{k}
=(\lambda{p}_i)p_i+q_i(\lambda\alpha^2q_i)-\dot{k}=\lambda\bigl(p_ip_i+\alpha^2q_iq_i\bigr)-\dot{k}={0}.
\end{eqnarray}
\noindent
A gauge-fixing condition must be accessible, starting from any initial configuration of the system.  So we would like to examine whether $\chi=0$ lies in the orbit generated by the constraint $C=0$.  Note, for constant $k$, we have $\dot{k}=0$ and $\dot{\chi}\neq{0}$ and the auxiliary condition cannot be preserved.  But for $k=k(t)$, this fixes the Lagrange multiplier $\lambda=\frac{\dot{k}(t)}{\vert\vec{p}\vert^2+\alpha^2\vert\vec{q}\vert^2}=\frac{\dot{k}(t)}{2\alpha^2r^2}$.  The meaning of $\lambda$ is unclear, especially since the Second Class approach furnished a well-defined time independently of any explicit $t$ dependence in $k(t)$.\par
\indent
Another flaw in the methodology of First Class constrained systems is when one wants to quantize the system via the so-called Wheeler--DeWitt equation \cite{CONSTRAINT1}.  Let us first examine the prescription, in the quantum theory, wherein Poisson brackets get promoted to $\frac{i}{\hbar}$ times commutators.  Since the relations are canonical, then the momentum acts on wavefunctions $\Psi(q_1,q_2,t)$ by differentiation $\hat{p}_i\Psi=\frac{\hbar}{i}\frac{\partial\Psi}{\partial{q}_i}$ and the physical states of the system must satisfy a Schrodinger equation $i\hbar\frac{\partial\Psi}{\partial{t}}=\hat{H}\Psi$.  On account of the First Class constraint $\hat{C}=0$ we get the following partial differential equation
\begin{eqnarray}
\label{FIRST4}
\hat{H}\Psi=\lambda\bigl(\hat{p}^2-\alpha^2\hat{q}^2\bigr)\Psi=0\longrightarrow\Bigl(\frac{\partial^2}{\partial{q_1}^2}+\frac{\partial^2}{\partial{q_2}^2}
+\frac{\alpha^2}{\hbar^2}(q_1^2+q_2^2)\Bigr)\Psi=0.
\end{eqnarray}
\noindent
This implies
$\frac{\partial\Psi}{\partial{t}}=0$, namely that the physical state $\Psi$ cannot evolve in time.  This is precisely what the procedure for Dirac quantization would prescribe, namely that the physical states be annihilated by the quantum constraints.  In General Relativity this leads to the problem of time in the analogous ``Wheeler--DeWitt" equation.\par
\indent
Constraints linear in momenta admit an interpretation of generating gauge transformations which preserve the configuration space polarization of the wavefunction for example as in Maxwell theory.  Since the wavefunctions $\Psi(q_1,q_2)$ are be defined on configuration and not phase space, then annihilation of the state by the constraint signifies its invariance under the transformation
\begin{eqnarray}
\Psi(q_i+\delta_{Gauge}{q}_i,t)=e^{(i/\hbar)\delta\vec{q}\cdot{\vec{C}}}\Psi=\Psi(q_i,t)+\sum_i\delta_{Gauge}{q}_i\frac{\partial\Psi}{\partial{q}_i}.
\end{eqnarray}
\noindent
However $C=p_ip_i-\alpha^2q_iq_i=0$ is quadratic in momenta, hence $C$ generates transformations which mix coordinate space with momentum space variables (\ref{WEHAVEIT}).  So the analogous Wheeler--DeWitt equation (\ref{FIRST4}) does not have the same meaning as a quantum wavefunction invariant under a transformation generated by $\hat{C}=0$, since the polarization  $\Psi(q_1,q_2)$ cannot be preserved.\par
\indent
It is interesting to note that ``First Class Dirac Quantization" procedure of having the constraint annihilate physical states,
 \begin{eqnarray}
\label{STILLHOLDS}
 {\hat C}\Psi(r,\varphi)=\frac{1}{2}({\hat p}^2_r +\frac {{\hat p}^2_\varphi}{r^2 }-\alpha^2r^2)\Psi(r,\varphi) =0,
  \end{eqnarray}
 can yield a correspondence if we were to replace ${\hat p}_r$ by $k/r$, it would yield the restriction $r=\Bigl(\frac{k^2+p^2_{\varphi}}{\alpha^2}\Bigr)^{1/4}$;
  and the results would correspond roughly to those presented in the earlier Second Class analysis. However,  ${\hat p}_r  \rightarrow k/r$ is consistent with the \textit{Dirac} bracket, and not \textit{Poisson} bracket, to commutator rule; and this rule cannot be deduced {\it a priori} within the context of First Class quantization without first turning the system into Second Class. Furthermore the Dirac bracket between $r$ and $\varphi$ does not vanish, so quantum states cannot be a function of both these variables. In fact this highlights another complication for First Class quantization: in a theory with $K$ First Class constraints, the reduced phase space will only be $2N-2K$ dimensional, so at most $(N-K)$, and not $N$ variables, can play the role of commuting configuration variables. It is far too optimistic in the First Class scheme to assume the $K$ constraints annihilating the wave function (which is still assumed to depend upon the original commuting $N$ configuration variables) will naturally reduce the system to one with $(N-K)$ {\it commuting configuration variables}, whereas examples show that some of the configuration variables may in fact be reduced into non-commuting variables.  As far as we know, in contradistinction to the Maskawa-Nakajima theorem for Second Class systems,
the reduction to a corresponding $2(N-K)$-dimensional classical phase space is not guaranteed by any theorem in First Class Dirac Quantization.

\section{8. Discussion and further remarks}

This work has illustrated the vast differences in the interpretation of totally constrained systems as seen from a Second Class versus First Class perspective.  The main motivation was to highlight some of the main problems which arise in General Relativity and to provide a resolution within the simpler context of a toy model sharing some of the relevant features in common.  We have argued that nature should exhibit a preference for Second Class over First Class systems with a view toward preservation of an even dimensional phase space.\par
\indent
There are several points which should be reiterated.  (i) In the case of unconstrained systems, then Poisson brackets and Dirac brakets coincide $\{A,B\}_{P.B}=\{A,B\}_{D.B.}$, and the standard quantization procedures carry through.  But where constraints exist, then they should be supplemented with auxiliary conditions to make the system Second Class.  The annihilation of physical states by quantum constraints  still holds, for example as in (\ref{STILLHOLDS}), upon promotion of \textit{Dirac} brackets, 
not \textit{Poisson} brackets, to commutators.  Unlike for First Class systems wherein the constraint fixes the state through a differential equation, for Second Class systems $\hat{\Phi}_I\vert\psi\rangle_{Phys}=0$  is trivially satisfied, and a Physical state $\vert\psi\rangle_{Phys}$ can be fixed only through unitary Hamiltonian evolution.  (ii) All phase space functions $A$ Dirac-commute with the constraints $\{A,\Phi_I\}_{D.B.}=0~\forall{A},\Phi_I$ by construction, hence are also observables in the quantum sense $ [\hat{A},\hat{\Phi}_I]=0$.  Nevertheless, for example transverse fields $(A^T_i(x),E^T_i(x))$, which form the reduced phase space for Maxwell theory are also observables in the First Class sense in that they are gauge-invariant.  (iii) While First Class constraints generate transformations of the phase space variables, this is not time evolution, as shown in all examples.  Time evolution can only arise due to a physical Hamiltonian $H_{Phys}$.  (vi) In the formulation of Second Class systems it may not be possible to satisfy $\hbox{det}\{\Phi_I,\Phi_J\}_{P.B.}\neq{0}$ on the whole phase space.  The reduced phase space can be defined such as to exclude such points when they form a set of measure zero.  For the Klauder problem the origin $(x^i,p_i)=(0,0)$ must be excluded.  But such exclusions are a small price to pay for access to the remaining part of phase space.  In the case of the Maxwell field, analogous to gravity \cite{SOOITAYU} the condition can be satisfied globally.  (iv) The Klauder toy problem suggests a strategy for the study of physical systems.\par 
\indent 
 One could envision the process of this article in reverse.  The Klauder problem can be seen as the particle in a plane wherein the radial momentum is placed upon the same footing as angular momentum (as the condition $k=rp_r$ is dimensionally consistent with the interpretation of $k$ as an angular momentum).  Then the problem becomes that of a particle on a circle, which can be solved exactly.  Using this as the starting point, then one may enlarge the phase space by introducing a radial degree of freedom $r$, along with its conugate momentum $p_r$ to preserve an even-dimensional phase space.  Then one constructs a constraint with auxiliary condition that can be invertibly mapped to these additional variables, and one has the Klauder problem and its variations for that system.  On a final note, one may analyse the effect wherein $k=k(t)$ is no longer a numerical constant and has explicit time dependence.  In this case we no longer have circular orbits as in (\ref{WEGET}), (\ref{WEGET11}), (\ref{EXPECTATION}).  Additionally, we now have a time-dependent Hamiltonian $U(\hat{p}_{\varphi},t)$ resulting in time-ordered unitary evolution of the state $\psi(\varphi,t)=U(t,t_0)\psi(\varphi,t_0)$ with
\begin{eqnarray}
\label{COMMUTATOR222}
U(t,t_0)={\cal T}\biggl[\hbox{exp}\Bigl[-\frac{i}{\hbar}\int^t_{t_0}U(\hat{p}_{\varphi},t')dt'\Bigr]\biggr]\nonumber\\
=I-\frac{i}{\hbar}\int^t_{t_0}dt_1U(\hat{p}_{\varphi},t_1)
+\Big(\frac{-i}{\hbar}\Bigr)^2\int^t_{t_0}dt_2\int^{t_2}_{t_0}dt_1\int^{t_2}_{t_0}dt_1(\hat{p}_{\varphi},t_2)(\hat{p}_{\varphi},t_1)+\dots\\
+\Big(\frac{-i}{\hbar}\Bigr)^n\int^t_{t_0}dt_n\int^{t_n}_{t_0}dt_{n-1}\int^{t_{n-1}}_{t_0}\dots\int^{t_2}_{t_0}dt_1(\hat{p}_{\varphi},t_n)(\hat{p}_{\varphi},t_{n-1})\dots(\hat{p}_{\varphi},t_1)+\dots
\end{eqnarray}
\noindent
Equation (\ref{COMMUTATOR222}) is analogous to the case for full gravity\cite{SOOYU23}, within the context of the relativistic point particle example considered; and the incorporation of transverse-traceless auxiliary conditions turning General Relativity into a Second Class system with a reduced Hamiltonian generating cosmic time translations has been addressed in Ref.\cite{SOOITAYU}.

From an observational standpoint, the theoretical results obtained through the techniques demonstrated in this articls should be true only true for experiments or observations that fulfill the associated constraints and auxiliary conditions.  From canonical equivalence of the phase space measure $\delta({\chi})\hbox{det}\{\chi,C\}_{P.B.}\delta(C)$, all suitable auxiliary conditions are quantum-mechanically equivalent.  This provides a vast amount of flexibility in setting up experiments and observations since one has an infinite number of auxiliary conditions to choose from.  So far, the best observational evidence in support of Second Class systems resides in the physicality of transverse electromagnetic fields, as well the detection of transverse- traceless perturbative gravitational waves.

\section{Acknowledgements}

The authors would like to thank John Klauder for posing this problem as a toy model for General Relativity, and for providing useful suggestions for this work.
This work has been supported in part by the  Ministry of Science and Technology (R.O.C.) under Grant No. MOST 110-2112-M-006-001.

\end{document}